\documentstyle[10pt,epsf,epsfig,dp_delphititle]{dp_delphi}
\makeindex
\def\DpPaperGroup{EP}
\def\DpPaperRef{2000-033}
\def\DpDate{7 June 2000}
\def\DpAuthors{DELPHI Collaboration}
\def\DpSubmit{(Phys. Lett. B485(2000)95)}
\def\DpTitle{{\bf Update of the search for charginos nearly mass-degenerate 
with the lightest neutralino }}
\def\DpComment{ }
\def\DpEMail{ }
\begin{document}
%%%%%%%%%%%%%%%%%%%%%%%%%% They are a problem with Coll.Sty ?
\makeatletter
%\input{dp_system:coll.sty}
% Collapse citation numbers to ranges.  Non-numeric and undefined labels
% are handled.  No sorting is done.  E.g., 1,3,2,3,4,5,foo,1,2,3,?,4,5
% gives 1,3,2-5,foo,1-3,?,4,5
\newcount\@tempcntc
\def\@citex[#1]#2{\if@filesw\immediate\write\@auxout{\string\citation{#2}}\fi
  \@tempcnta\z@\@tempcntb\m@ne\def\@citea{}\@cite{\@for\@citeb:=#2\do
    {\@ifundefined
       {b@\@citeb}{\@citeo\@tempcntb\m@ne\@citea\def\@citea{,}{\bf ?}\@warning
       {Citation `\@citeb' on page \thepage \space undefined}}%
    {\setbox\z@\hbox{\global\@tempcntc0\csname b@\@citeb\endcsname\relax}%
     \ifnum\@tempcntc=\z@ \@citeo\@tempcntb\m@ne
       \@citea\def\@citea{,}\hbox{\csname b@\@citeb\endcsname}%
     \else
      \advance\@tempcntb\@ne
      \ifnum\@tempcntb=\@tempcntc
      \else\advance\@tempcntb\m@ne\@citeo
      \@tempcnta\@tempcntc\@tempcntb\@tempcntc\fi\fi}}\@citeo}{#1}}
\def\@citeo{\ifnum\@tempcnta>\@tempcntb\else\@citea\def\@citea{,}%
  \ifnum\@tempcnta=\@tempcntb\the\@tempcnta\else
   {\advance\@tempcnta\@ne\ifnum\@tempcnta=\@tempcntb \else \def\@citea{--}\fi
    \advance\@tempcnta\m@ne\the\@tempcnta\@citea\the\@tempcntb}\fi\fi}
 
\makeatother
%%%%%%%%%%%%%%%%%%%%%%%%%% ??????????????????????????????????
% Generate the title page
\begin{titlepage}
\pagenumbering{roman}
\CERNpreprint{\DpPaperGroup}{\DpPaperRef} % Reference of the paper
\date{{\small\DpDate}} % Date of the paper
\title{\DpTitle} % Title of the paper
\address{\DpAuthors} % General name of the author(s)
\begin{shortabs} % Start the abstract
\noindent
%   abstract.tex
%
\noindent

The data collected by DELPHI in 1998 at the centre-of-mass energy of 189~GeV
have been used to update the search for charginos nearly mass-degenerate with
the lightest supersymmetric particle, which is assumed to be the lightest
neutralino. Mass differences below $\Delta M\!=\!3$~GeV/c$^2$ are considered. 
No excess of events with respect to the Standard Model expectation has been
observed, and exclusions in the plane of $\Delta M$ versus chargino mass are
given. The new $\Delta M$ independent lower limit on the mass of the chargino
is $62.4$~GeV/c$^2$ in the higgsino scenario (which includes the gaugino mass 
unification scenario), if all sfermions are heavier than the lightest chargino.
In the approximation of large sfermion masses the limit is $59.8$~GeV/c$^2$, 
independently of the field content.
\end{shortabs}
\vfill
\begin{center}
\DpSubmit \ \\ % Horrible hack to allow to have DpSubmit empty
\DpComment \ \\
\DpEMail \ \\
\end{center}
\vfill
\clearpage
\headsep 10.0pt
\addtolength{\textheight}{10mm}
\addtolength{\footskip}{-5mm}
\begingroup
% Commands to process the author names
%
\newcommand{\DpName}[2]{\hbox{#1$^{\ref{#2}}$},\hfill}
\newcommand{\DpNameTwo}[3]{\hbox{#1$^{\ref{#2},\ref{#3}}$},\hfill}
\newcommand{\DpNameThree}[4]{\hbox{#1$^{\ref{#2},\ref{#3},\ref{#4}}$},\hfill}
\newskip\Bigfill \Bigfill = 0pt plus 1000fill
\newcommand{\DpNameLast}[2]{\hbox{#1$^{\ref{#2}}$}\hspace{\Bigfill}}
%
%\small
\footnotesize
\noindent
\DpName{P.Abreu}{LIP}
\DpName{W.Adam}{VIENNA}
\DpName{T.Adye}{RAL}
\DpName{P.Adzic}{DEMOKRITOS}
\DpName{I.Ajinenko}{SERPUKHOV}
\DpName{Z.Albrecht}{KARLSRUHE}
\DpName{T.Alderweireld}{AIM}
\DpName{G.D.Alekseev}{JINR}
\DpName{R.Alemany}{VALENCIA}
\DpName{T.Allmendinger}{KARLSRUHE}
% ...
\normalsize
\endgroup
\titlefoot{Department of Physics and Astronomy, Iowa State
     University, Ames IA 50011-3160, USA
    \label{AMES}}
\titlefoot{Physics Department, Univ. Instelling Antwerpen,
     Universiteitsplein 1, B-2610 Antwerpen, Belgium \\
     \indent~~and IIHE, ULB-VUB,
     Pleinlaan 2, B-1050 Brussels, Belgium \\
     \indent~~and Facult\'e des Sciences,
     Univ. de l'Etat Mons, Av. Maistriau 19, B-7000 Mons, Belgium
    \label{AIM}}
\titlefoot{Physics Laboratory, University of Athens, Solonos Str.
     104, GR-10680 Athens, Greece
    \label{ATHENS}}
\titlefoot{Department of Physics, University of Bergen,
     All\'egaten 55, NO-5007 Bergen, Norway
    \label{BERGEN}}
\titlefoot{Dipartimento di Fisica, Universit\`a di Bologna and INFN,
     Via Irnerio 46, IT-40126 Bologna, Italy
    \label{BOLOGNA}}
\titlefoot{Centro Brasileiro de Pesquisas F\'{\i}sicas, rua Xavier Sigaud 150,
     BR-22290 Rio de Janeiro, Brazil \\
     \indent~~and Depto. de F\'{\i}sica, Pont. Univ. Cat\'olica,
     C.P. 38071 BR-22453 Rio de Janeiro, Brazil \\
     \indent~~and Inst. de F\'{\i}sica, Univ. Estadual do Rio de Janeiro,
     rua S\~{a}o Francisco Xavier 524, Rio de Janeiro, Brazil
    \label{BRASIL}}
\titlefoot{Comenius University, Faculty of Mathematics and Physics,
     Mlynska Dolina, SK-84215 Bratislava, Slovakia
    \label{BRATISLAVA}}
\titlefoot{Coll\`ege de France, Lab. de Physique Corpusculaire, IN2P3-CNRS,
     FR-75231 Paris Cedex 05, France
    \label{CDF}}
\titlefoot{CERN, CH-1211 Geneva 23, Switzerland
    \label{CERN}}
\titlefoot{Institut de Recherches Subatomiques, IN2P3 - CNRS/ULP - BP20,
     FR-67037 Strasbourg Cedex, France
    \label{CRN}}
\titlefoot{Now at DESY-Zeuthen, Platanenallee 6, D-15735 Zeuthen, Germany
    \label{DESY}}
\titlefoot{Institute of Nuclear Physics, N.C.S.R. Demokritos,
     P.O. Box 60228, GR-15310 Athens, Greece
    \label{DEMOKRITOS}}
\titlefoot{FZU, Inst. of Phys. of the C.A.S. High Energy Physics Division,
     Na Slovance 2, CZ-180 40, Praha 8, Czech Republic
    \label{FZU}}
\titlefoot{Dipartimento di Fisica, Universit\`a di Genova and INFN,
     Via Dodecaneso 33, IT-16146 Genova, Italy
    \label{GENOVA}}
\titlefoot{Institut des Sciences Nucl\'eaires, IN2P3-CNRS, Universit\'e
     de Grenoble 1, FR-38026 Grenoble Cedex, France
    \label{GRENOBLE}}
\titlefoot{Helsinki Institute of Physics, HIP,
     P.O. Box 9, FI-00014 Helsinki, Finland
    \label{HELSINKI}}
\titlefoot{Joint Institute for Nuclear Research, Dubna, Head Post
     Office, P.O. Box 79, RU-101 000 Moscow, Russian Federation
    \label{JINR}}
\titlefoot{Institut f\"ur Experimentelle Kernphysik,
     Universit\"at Karlsruhe, Postfach 6980, DE-76128 Karlsruhe,
     Germany
    \label{KARLSRUHE}}
\titlefoot{Institute of Nuclear Physics and University of Mining and Metalurgy,
     Ul. Kawiory 26a, PL-30055 Krakow, Poland
    \label{KRAKOW}}
\titlefoot{Universit\'e de Paris-Sud, Lab. de l'Acc\'el\'erateur
     Lin\'eaire, IN2P3-CNRS, B\^{a}t. 200, FR-91405 Orsay Cedex, France
    \label{LAL}}
\titlefoot{School of Physics and Chemistry, University of Lancaster,
     Lancaster LA1 4YB, UK
    \label{LANCASTER}}
\titlefoot{LIP, IST, FCUL - Av. Elias Garcia, 14-$1^{o}$,
     PT-1000 Lisboa Codex, Portugal
    \label{LIP}}
\titlefoot{Department of Physics, University of Liverpool, P.O.
     Box 147, Liverpool L69 3BX, UK
    \label{LIVERPOOL}}
\titlefoot{LPNHE, IN2P3-CNRS, Univ.~Paris VI et VII, Tour 33 (RdC),
     4 place Jussieu, FR-75252 Paris Cedex 05, France
    \label{LPNHE}}
\titlefoot{Department of Physics, University of Lund,
     S\"olvegatan 14, SE-223 63 Lund, Sweden
    \label{LUND}}
\titlefoot{Universit\'e Claude Bernard de Lyon, IPNL, IN2P3-CNRS,
     FR-69622 Villeurbanne Cedex, France
    \label{LYON}}
\titlefoot{Univ. d'Aix - Marseille II - CPP, IN2P3-CNRS,
     FR-13288 Marseille Cedex 09, France
    \label{MARSEILLE}}
\titlefoot{Dipartimento di Fisica, Universit\`a di Milano and INFN-MILANO,
     Via Celoria 16, IT-20133 Milan, Italy
    \label{MILANO}}
\titlefoot{Dipartimento di Fisica, Univ. di Milano-Bicocca and
     INFN-MILANO, Piazza delle Scienze 2, IT-20126 Milan, Italy
    \label{MILANO2}}
\titlefoot{Niels Bohr Institute, Blegdamsvej 17,
     DK-2100 Copenhagen {\O}, Denmark
    \label{NBI}}
\titlefoot{IPNP of MFF, Charles Univ., Areal MFF,
     V Holesovickach 2, CZ-180 00, Praha 8, Czech Republic
    \label{NC}}
\titlefoot{NIKHEF, Postbus 41882, NL-1009 DB
     Amsterdam, The Netherlands
    \label{NIKHEF}}
\titlefoot{National Technical University, Physics Department,
     Zografou Campus, GR-15773 Athens, Greece
    \label{NTU-ATHENS}}
\titlefoot{Physics Department, University of Oslo, Blindern,
     NO-1000 Oslo 3, Norway
    \label{OSLO}}
\titlefoot{Dpto. Fisica, Univ. Oviedo, Avda. Calvo Sotelo
     s/n, ES-33007 Oviedo, Spain
    \label{OVIEDO}}
\titlefoot{Department of Physics, University of Oxford,
     Keble Road, Oxford OX1 3RH, UK
    \label{OXFORD}}
\titlefoot{Dipartimento di Fisica, Universit\`a di Padova and
     INFN, Via Marzolo 8, IT-35131 Padua, Italy
    \label{PADOVA}}
\titlefoot{Rutherford Appleton Laboratory, Chilton, Didcot
     OX11 OQX, UK
    \label{RAL}}
\titlefoot{Dipartimento di Fisica, Universit\`a di Roma II and
     INFN, Tor Vergata, IT-00173 Rome, Italy
    \label{ROMA2}}
\titlefoot{Dipartimento di Fisica, Universit\`a di Roma III and
     INFN, Via della Vasca Navale 84, IT-00146 Rome, Italy
    \label{ROMA3}}
\titlefoot{DAPNIA/Service de Physique des Particules,
     CEA-Saclay, FR-91191 Gif-sur-Yvette Cedex, France
    \label{SACLAY}}
\titlefoot{Instituto de Fisica de Cantabria (CSIC-UC), Avda.
     los Castros s/n, ES-39006 Santander, Spain
    \label{SANTANDER}}
\titlefoot{Dipartimento di Fisica, Universit\`a degli Studi di Roma
     La Sapienza, Piazzale Aldo Moro 2, IT-00185 Rome, Italy
    \label{SAPIENZA}}
\titlefoot{Inst. for High Energy Physics, Serpukov
     P.O. Box 35, Protvino, (Moscow Region), Russian Federation
    \label{SERPUKHOV}}
\titlefoot{J. Stefan Institute, Jamova 39, SI-1000 Ljubljana, Slovenia
     and Laboratory for Astroparticle Physics,\\
     \indent~~Nova Gorica Polytechnic, Kostanjeviska 16a, SI-5000 Nova Gorica, Slovenia, \\
     \indent~~and Department of Physics, University of Ljubljana,
     SI-1000 Ljubljana, Slovenia
    \label{SLOVENIJA}}
\titlefoot{Fysikum, Stockholm University,
     Box 6730, SE-113 85 Stockholm, Sweden
    \label{STOCKHOLM}}
\titlefoot{Dipartimento di Fisica Sperimentale, Universit\`a di
     Torino and INFN, Via P. Giuria 1, IT-10125 Turin, Italy
    \label{TORINO}}
\titlefoot{Dipartimento di Fisica, Universit\`a di Trieste and
     INFN, Via A. Valerio 2, IT-34127 Trieste, Italy \\
     \indent~~and Istituto di Fisica, Universit\`a di Udine,
     IT-33100 Udine, Italy
    \label{TU}}
\titlefoot{Univ. Federal do Rio de Janeiro, C.P. 68528
     Cidade Univ., Ilha do Fund\~ao
     BR-21945-970 Rio de Janeiro, Brazil
    \label{UFRJ}}
\titlefoot{Department of Radiation Sciences, University of
     Uppsala, P.O. Box 535, SE-751 21 Uppsala, Sweden
    \label{UPPSALA}}
\titlefoot{IFIC, Valencia-CSIC, and D.F.A.M.N., U. de Valencia,
     Avda. Dr. Moliner 50, ES-46100 Burjassot (Valencia), Spain
    \label{VALENCIA}}
\titlefoot{Institut f\"ur Hochenergiephysik, \"Osterr. Akad.
     d. Wissensch., Nikolsdorfergasse 18, AT-1050 Vienna, Austria
    \label{VIENNA}}
\titlefoot{Inst. Nuclear Studies and University of Warsaw, Ul.
     Hoza 69, PL-00681 Warsaw, Poland
    \label{WARSZAWA}}
\titlefoot{Fachbereich Physik, University of Wuppertal, Postfach
     100 127, DE-42097 Wuppertal, Germany
    \label{WUPPERTAL}}
\addtolength{\textheight}{-10mm}
\addtolength{\footskip}{5mm}
\clearpage
\headsep 30.0pt
\end{titlepage}
%%%%%%%%%%%%%%%%%%%%%%%%%
%
% Change for the document body
%%\pagestyle{heading} % for page numbering
\pagenumbering{arabic} % page numbering in number
\setcounter{footnote}{0} %
\large
%\linenumbers %%%CD
%   document.tex
%

%===================> ADD here your LATEX definitions
\def\deg{^\circ}
\def\mmin{\mathrm{min}}

\def\cguz{\tilde\chi_1^0}
\def\cgdz{\tilde\chi_2^0}
\def\cgpm{\tilde\chi_1^{\pm}}
\def\cgdpm{\tilde\chi_2^{\pm}}
\def\cgup{\tilde\chi_1^+}
\def\cgum{\tilde\chi_1^-}
\def\snu{\tilde\nu}
\def\mcguz{M_{\cguz}}
\def\mcgup{M_{\cgup}}
\def\mcgpm{M_{\cgup}}
\def\msnu{M_{\snu}}

\def\gamgam{\gamma\gamma}

\def\DMP{\Delta M}
\def\eff{\varepsilon}
\def\nup{N_{\rm up}}

\def\gevc2{GeV/c$^2$}
\def\mevc2{MeV/c$^2$}

\def\ecms{E_{\mathrm{cms}}}
\def\evis{E_{\mathrm{vis}}}
\def\mopp{M_{\mathrm{opp}}}
\def\etgam{E_T^{\gamma}}
\def\etvis{E_T^{\mathrm{vis}}}
\def\ptmiss{P_T^{\mathrm{miss}}}

%===================> MAIN part

\section{Introduction}

This paper updates the results of the search for charginos ($\cgpm$) nearly 
mass-degenerate with the lightest neutralino ($\cguz$) reported in 
Ref.~\cite{paper214}, with the data collected by DELPHI in 1998 at the 
centre-of-mass energy of 189~GeV.

The experimental techniques used depend on the mass difference $\DMP$ between 
the chargino and the lightest neutralino (assumed to be the Lightest
Supersymmetric Particle, LSP), as described in Ref.~\cite{paper214}.
When $\DMP$ is below the mass of the pion, the chargino lifetime is typically
long enough to let it pass through the entire detector before decaying. 
This range of $\DMP$ can be covered by the search for long-lived heavy charged
particles. 
For $\DMP$ of few hundred \mevc2  the $\cgpm$ can decay inside the main
tracking devices of DELPHI. Therefore, a search for secondary vertices or 
kinks can be used to explore this region.
With increasing mass difference, the mean lifetime falls and it becomes
difficult to distinguish the position of the $\cgpm$ decay vertex from the
initial interaction point.
In this case, the tagging of a high energy Initial State Radiation (ISR) photon
can help in exploring the $\DMP$ region between a few hundred \mevc2 and 
3~\gevc2.

Compared to Ref.~\cite{paper214}, the search using a tagged ISR photon has 
been improved with the use of an additional cut and a wider range of mass 
differences between the chargino and the lightest neutralino explored.
Moreover, the selection cuts were optimized for each point in the plane
$(\mcgup,\DMP)$, depending on the kinematics of the signal in that point.
This significantly improved the sensitivity in a region of the space of the SUSY
parameters that will probably never be covered by the searches at hadron
machines~\cite{newgunion}.

All the new data have been combined with the samples already used in
Ref.~\cite{paper214}. In the search which uses ISR, all the old data-sets 
have been re-analysed according to the new prescriptions.

Three SUSY scenarios were considered,
depending on the values of the SU(2) gaugino mass $M_2$, the U(1) gaugino
mass $M_1$, the Higgs mixing parameter $\mu$, and the mass of the
sneutrino $\msnu$:
\begin{enumerate}
 \item{$M_{1,2} \gg |\mu|$                            (higgsino-like);}
 \item{$|\mu| \gg M_1 \ge M_2$ and heavy sneutrino    (gaugino-like);}
 \item{$|\mu| \gg M_1 \ge M_2$ and light sneutrino    (gaugino-like).}
\end{enumerate}

Gauginos couple to $\snu$, thus heavy and light sneutrinos define two
phenomenologically different gaugino scenarios, with different cross-sections
(because of the possible chargino production through $\snu$ exchange in the
$t$-channel), lifetimes and branching ratios. 
In the following, in the heavy 
sneutrino scenario $\msnu>500$~\gevc2 is assumed. In all other cases the
assumption is $\msnu>\mcgpm$.
%In the light sneutrino scenario, a lower bound $\msnu>\mcgpm + 500$~\mevc2 is
%considered as an example in the searches for long-lived charginos: lifetimes
%in fact become quickly extremely short when sneutrinos approach from above the
%mass of the chargino. However, the results obtained in the search which uses
%the ISR tag are valid for any $\msnu>\mcgpm$.

Also the charged sfermions couple to the gauginos, and if light they can
modify the lifetimes and the branching ratios considered. In the following,
the heavy charged sfermions approximation will be used for the two
gaugino scenarios, while for the higgsino it is enough to consider 
$M_{\tilde f_i} > \mcgup$ for all sfermions.

A charged gaugino (cases 2 and 3) can get a mass 
$\mcgpm \leq \mcguz + 1$~\gevc2 only if the constraint of gaugino mass
unification at the GUT scale, implying the electroweak scale relation
$M_1 \sim 1/2 \, M_2$, is released. Several interesting scenarios 
without gaugino mass unification or with near mass-degeneracy between the
lightest supersymmetric particles have been proposed 
\cite{cdg,th1,thomas,th2,th3}, and all of them can be studied by using the
techniques reported in the present paper.

\section{Data samples and event generators}
\label{sec:data}

The DELPHI detector is described in \cite{delphidet}.
The integrated luminosity collected by DELPHI at 189~GeV was approximately
158~pb$^{-1}$, out of which 155.3~pb$^{-1}$ were used in the searches for
long-lived particles and 152.9~pb$^{-1}$ in the search for soft particles
accompanied by an ISR photon. 

{\tt SUSYGEN}~\cite{susygen} was used to generate all signal samples and to
calculate cross-sections. The decay modes of the chargino when $\DMP < 2$~\gevc2
were modelled using the computation of~\cite{cdg}, while the widths given by
{\tt SUSYGEN} were used for $\DMP \ge 2$~\gevc2. About 325000 events were 
generated, in correspondence of six different chargino masses and seven
different $\DMP$'s.

The background process $e^+e^- \to q\bar q$ ($n\gamma$) was generated with
{\tt PYTHIA 5.7}~\cite{JETSET}, while {\tt DYMU3}~\cite{DYMU3} and
{\tt KORALZ 4.2}~\cite{KORALZ} were used for $\mu ^+\mu ^-(\gamma)$ and
$\tau^+\tau^-(\gamma)$, respectively.
Processes leading to four-fermion final states, $(Z/\gamma)^*(Z\gamma)^*$,
$W^+W^-$, $We\nu$ and $Zee$, were generated using
{\tt EXCALIBUR}~\cite{EXCALIBUR} and {\tt GRC4F}~\cite{GRACE}.

Two-photon interactions leading to hadronic final states were generated using
{\tt TWOGAM}~\cite{TWOGAM}, including the VDM, QCD and QPM components.
The generators of~\cite{BDK} were used for the leptonic final states.
As in the previous analysis~\cite{paper214}, one had to deal with the fact
that part of the two-photon background was not simulated, because of 
phase-space cuts applied during the event generation. For instance, in the
simulation at 189 GeV the $e^+e^-$ final state had no ISR; in the hadronic
samples, the mass of the two photon system was required to be 
$M_{\gamgam}>3$~\gevc2, and at the same time there had to be at least one
charged particle with $p_T>1.2$~GeV/c.

All generated signal and background events were passed through a detailed
simulation of the DELPHI detector~\cite{delsim} and then processed
with the same reconstruction and analysis programs as real data events.
The number of simulated events from different background processes was
several times the number of real events recorded.

\section{Search for long-lived charginos}

\subsection{Search for heavy stable charged particles}

The results of the search for heavy stable charged particles at 189 GeV are
described in Ref.~\cite{stable189}, where all the details on the techniques
used and on the efficiency can be found. The efficiencies for selecting heavy
stable particles presented there were then convoluted with the
expected distribution of the decay length of the chargino in a given scenario,
in order to derive an event selection efficiency for long-lived charginos
as a function of their mass and lifetime.
One event was selected in the data, while $1.02 \pm 0.13$ events were expected
from Standard Model (SM) processes.

\subsection{Search for decay vertices inside the detector}

To search for chargino decays inside the sensitive detector volume the same
selection as in Ref.~\cite{paper214} was used. No events remained
in the data collected at 189 GeV. The number of background events expected from
SM processes was $0.87\pm 0.65$ ($0.63$ from Bhabha scattering, $0.12$
from $e^+e^- \to \mu^+\mu^-$ and $0.12$ from 
$e^+e^- \to \tau^+\tau^-$).

\begin{figure}[bht]
\centerline{
\epsfxsize=10.cm\epsffile{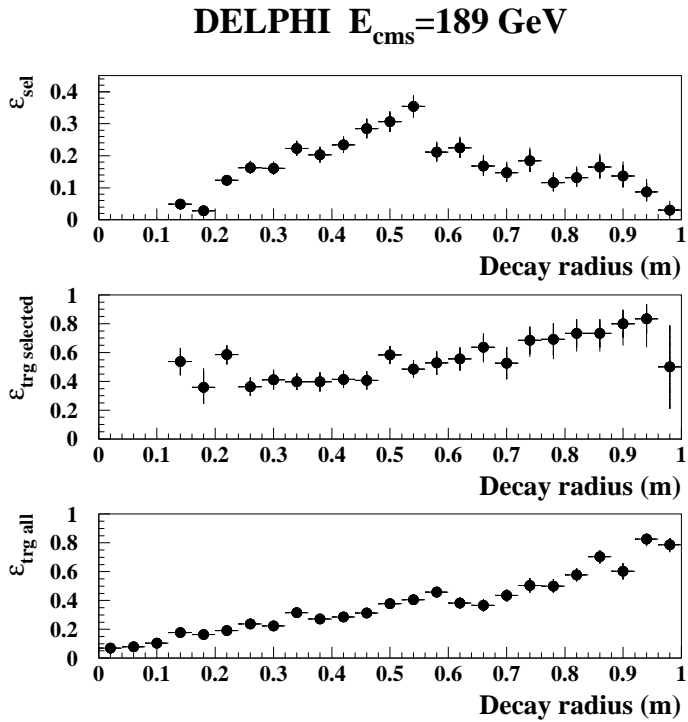} }
\caption[]{ Top: efficiency for selecting a single 70~\gevc2 chargino in
  the search for displaced decay vertices (kinks), as function of its decay
  radius in DELPHI. Middle: trigger efficiency for charginos selected
  with the offline criteria. Below: trigger efficiency for all 70~\gevc2
  charginos, whether or not they are selected.
                     }
\label{fig:kinkeff}
\end{figure}

The efficiencies for the signal have been estimated using simulated samples of
charginos with different decay lengths. Figure \ref{fig:kinkeff} shows, as
an example, efficiencies as functions of the decay radius for a chargino mass 
of $70$~\gevc2. The first plot shows the efficiency for selecting a
single chargino. 
The efficiency first increases with the decay radius since longer chargino
tracks are better reconstructed. For even larger radii
the efficiency decreases due to the
poorer reconstruction of the low momentum decay products. The vertex
reconstruction reaches the maximum of the efficiency in the middle of the TPC.
The second plot displays the trigger efficiency for the charginos passing the
selection criteria. The third plot shows the trigger efficiency for all 
charginos of that mass, whether or not they were selected. 
Trigger efficiencies were estimated by using a Monte Carlo simulation of the
performances of the relevant trigger components.

\subsection{Results in the search for long-lived charginos}

In the absence of evidence for a signal in any of the searches for long-lived
charginos at 189 GeV, the results of the two methods can be combined,
as explained in Ref.~\cite{paper214}. Again, these results can be further 
combined with the outcomes of the search at lower centre-of-mass energies, also
described in Ref.~\cite{paper214}. The regions excluded, with a confidence 
level (CL) of at least 95\%, by such a combination of searches for long-lived 
charginos in the plane $(\mcgup,\DMP)$ will be shown in figure \ref{fig:limit}.
No limit is derived in the gaugino scenario with light $\snu$, since the
lifetime limit cancels out when $\msnu$ approaches $\mcgup$.
%In the light $\snu$ scenario the special case $\msnu=\mcgup + 500$~\mevc2 is
%considered as an example, since the lifetime limit cancel out when $\msnu$
%approaches $\mcgup$.

\section{Search for charginos with ISR photons}

With respect to the analysis described in Ref.~\cite{paper214} a new variable
was taken into account to better discriminate between nearly mass-degenerate 
charginos and the dominant two-photon background. This variable is the ratio
between the missing transverse momentum ($\ptmiss$) and the visible transverse
energy ($\etvis$) in the event. 
Another improvement in the analysis at 189~GeV is that for $\DMP<1$~\gevc2
the requirement of at least two charged tracks consistent with coming
from a common primary vertex was removed. This increased the efficiency for 
events with charginos decaying up to a few~cm from the interaction point.

In summary, after a common preselection, which remained unchanged with respect
to Ref.~\cite{paper214}, the cuts applied to the data and to the
simulated signal and background samples, as functions of $\mcgup$ and $\DMP$,
were the following ($\ecms$ is the centre-of-mass energy):

\begin{itemize}
\item{There must be at least two and at most six good charged particles and,
 in any case, no more than ten tracks in the event. Tracks reconstructed in the
 tracking devices of DELPHI are taken as good charged particles if they have
 a momentum above 100~MeV/c, measured with $\delta p / p < 100\%$, and
 an impact parameter below 4~cm in the azimuthal plane and below 10~cm in the
 longitudinal plane. }
\item{The transverse energy of the ISR photon was required to be greater than
 $(\etgam)^{\mmin}$, where $(\etgam)^{\mmin} \simeq 0.03\cdot\ecms$.}
\item{The mass recoiling against the photon must be above $2\mcgup-\delta M$,
 where the term $\delta M$ takes into account the energy resolution in the
 electromagnetic calorimeters.}
\item{The photon had to be isolated by at least $30\deg$ with respect
 to any other charged or neutral particle in the event.}
\item{The sum of the energies of the particles emitted within $30\deg$ to the
 beam axis ($E_{30}$) was required to be less than 25\% of the total visible
 energy. If the photon was inside this angular region its energy was included
 neither in $E_{30}$, nor in the visible energy.}
\item{If the ISR photon candidate was detected in the very forward DELPHI
 calorimeter (STIC), it must not be correlated with a signal in the
 scintillators placed in front of STIC.}
\item{(In the data collected since 1997) if the ISR photon candidate was at an
 angle between $10\deg$ and $25\deg$ with respect to the beam direction, the
 region where the Time Projection Chamber (TPC) cannot be used in the tracking,
 it must not be correlated with hits in the Silicon Tracker.}
\item{$(\evis-E_{\gamma})/ \ecms$ must be below a kinematical threshold
 which depends on $\DMP$ and on $\mcgup$ (and in any case below  $6\%$).}
\item{$\ptmiss / \etvis$ must be above 0.40 if $\DMP>300$~\mevc2, and 
 above 0.75 for smaller $\DMP$'s.}
\item{If $\DMP>1$~\gevc2, at least two charged particles in the event must be
 consistent with coming from the interaction vertex.}
\end{itemize}

\begin{figure}[tbh]
\centerline{
\epsfxsize=12cm\epsffile{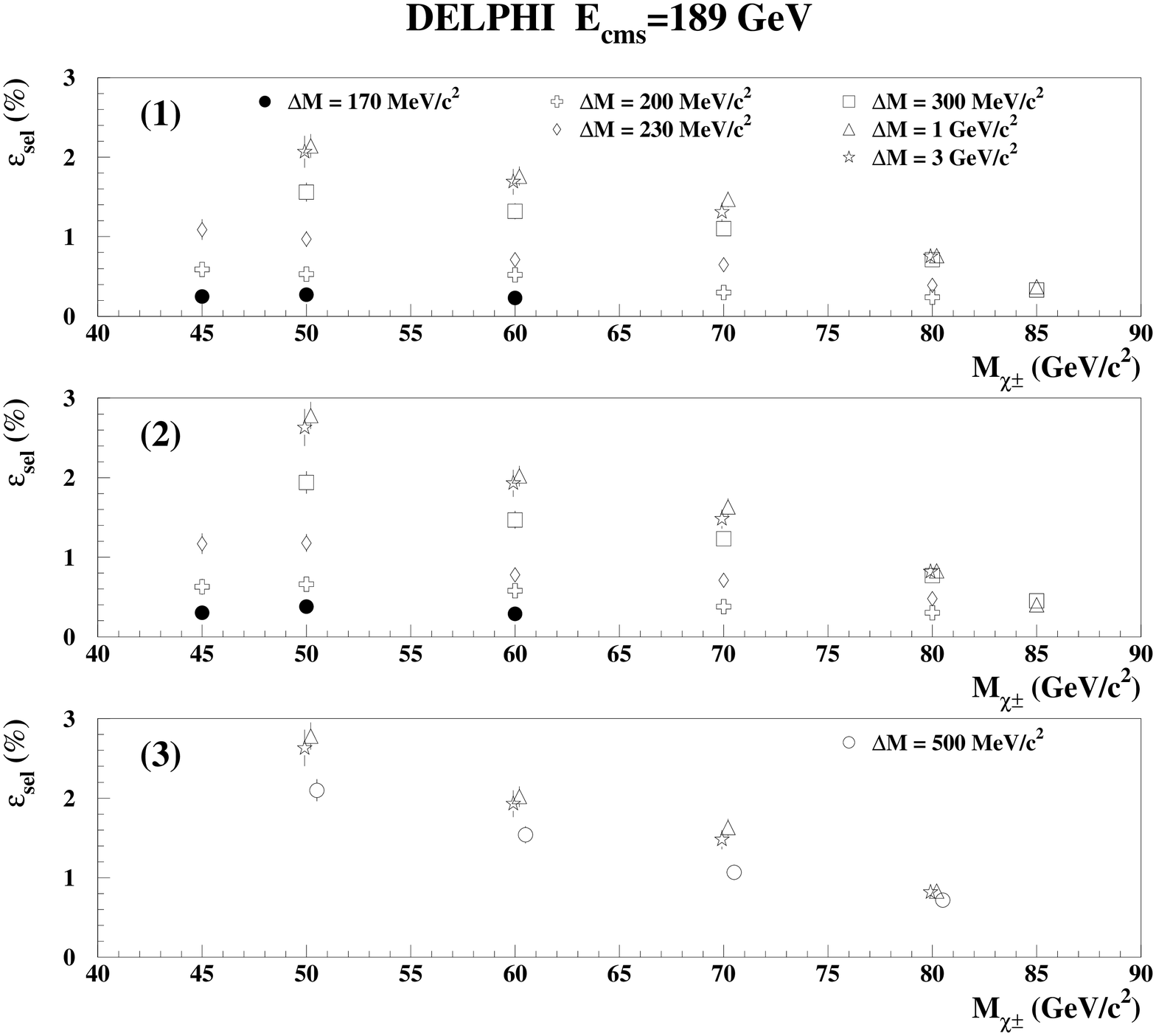} }
\caption[]{ Selection efficiencies in the search with ISR for charged 
 higgsinos (1), charged gauginos in case of heavy sneutrinos (2) and charged
 gauginos in case of light sneutrinos (3) at the centre-of-mass energy of
 189~GeV, as functions of their mass and of the mass difference with the
 lightest neutralino. 
             }
\label{fig:isreff}
\end{figure}

Figure \ref{fig:isreff} shows the efficiencies for the signal of nearly 
mass-degenerate charginos, computed with the fully simulated samples at
189~GeV, for some values of $\mcgup$ and $\DMP$. The difference between the
efficiencies in the higgsino and in the gaugino scenario with heavy $\snu$
are due to the different ISR energy spectrum. The efficiencies for gauginos
in case of light sneutrinos get smaller at the lowest $\DMP$ because of the
larger fraction of missing energy in the $\cgup \to l^+\nu\cguz$ decay.
Mass differences below 500~\mevc2 were not simulated in the light sneutrino
scenario. 

The overall trigger efficiency for these samples depend on $\mcgup$ and $\DMP$
and it was found to always be above $78\%$. It was obtained as the logical OR
of the single photon trigger efficiency~\cite{singleg} and the trigger 
efficiency for low momentum tracks. Both trigger efficiencies were measured by
using the data collected by DELPHI during the same period of data acquisition.

\subsection{Results in the search for charginos with ISR photons}

\begin{figure}[tbh]
\centerline{
\epsfxsize=10.0cm\epsffile{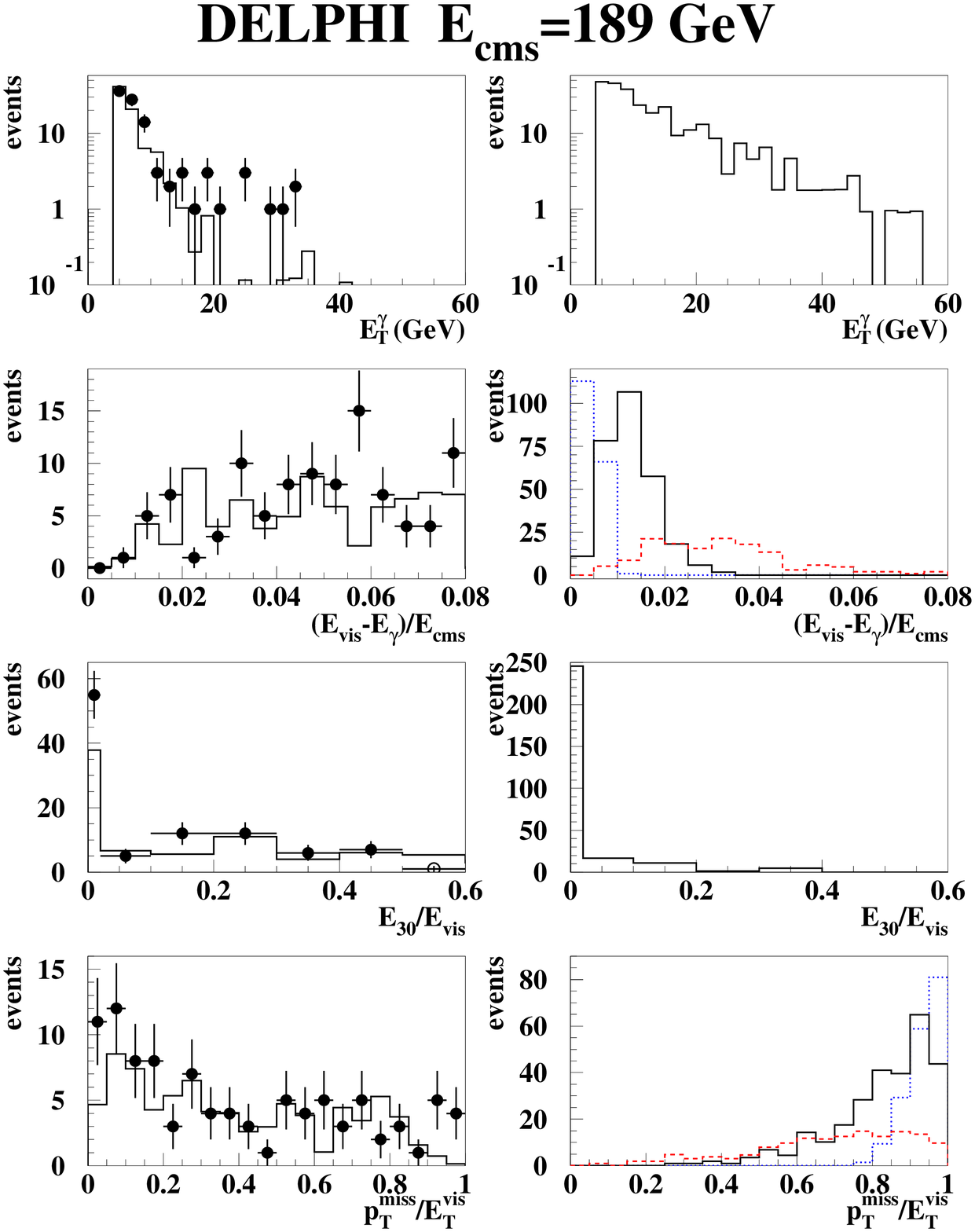} }
\caption[]{ Some of the variables used in the selection at $189$ GeV.
    In the left plots the data (dots) are compared with the SM expectations.
    On the right, as an example, the corresponding distributions (with
    arbitrary normalisation) are shown for the signal with $\mcgup = 60$~\gevc2
    and $\DMP =1$~\gevc2. In the plot of the visible energy (second row) and of
    the ratio $\ptmiss/\etvis$ (last row) three different mass splittings are
    shown for the signal: dotted, $\DMP =0.3$~\gevc2; solid line,
    $\DMP =1$~\gevc2; dashed, $\DMP =3$~\gevc2.                     }
\label{fig:bck189}
\end{figure}

In spite of the limitation due to the cuts applied at the generation on the 
two-photon samples, described in section~\ref{sec:data}, the agreement 
between the data and the simulation is reasonable after the preselection and
the removal of all events where the candidate ISR photon was compatible with
being a charged particle in the forward region. A comparison can be seen in 
figure~\ref{fig:bck189}, where the distributions of the transverse energy of 
the photon, of the visible energy besides the ISR photon, of the visible 
energy within $30\deg$ to the beam axis and of the ratio $\ptmiss/\etvis$ are
shown for the data (dots), for the sum of the SM backgrounds (left histograms)
and for the signal sample with $\mcgup=60$~\gevc2 (histograms on the right).

There is indeed some excess of data over simulated background in signal-like
regions in the plots of the transverse energy of the photon and of the ratio
$\ptmiss/\etvis$, although of little statistical significance.
Once all the selections cuts were applied, however, only one event at 189~GeV 
with a photon with $\etgam > 20$~GeV remained, where $0.36\pm0.20$ were
expected.

%able \ref{tab:isrcand} gives the number of events observed and the expected 
Table~1 gives the number of events observed and the expected 
background in the search at 189 GeV and after the re-analysis of all the data
collected at the lower centre-of-mass energies. The logical OR of the 
selections, for all masses and $\DMP$ considered, was used.

\begin{table}[thbp]
{ \small
\begin{center}
\begin{tabular}{|c||c|c|c||c|}
 \hline
      \makebox[5em] {\bf  Data                  }  &
      \makebox[7em] {\boldmath \bf  $Z^0\gamma \to \tau^+\tau^-$     }  &
      \makebox[7em] {\boldmath \bf  $\gamgam \to$~hadr.              }  &
      \makebox[7em] {\boldmath \bf  $\gamgam \to \tau^+\tau^-$       }  &
      \makebox[8em] {\boldmath \bf $\Sigma$ bckg  }  \\
 \hline
 \hline
 \multicolumn{5}{|c|} {
 {\boldmath \bf $\ecms=130$/$136$ GeV} \,\,\,\,\,\,(~$\int {\cal L} = 11.7$ pb$^{-1}$) } \\
   $0$  & - & $0.10 \pm 0.10$  & - &  $0.10 \pm 0.10$ \\
 \hline
 \hline
 \multicolumn{5}{|c|} {
 {\boldmath \bf $\ecms=161$ GeV} \,\,\,\,\,\,(~$\int {\cal L} = 9.7$ pb$^{-1}$) } \\
   $0$  & - & $0.67 \pm 0.32$  & $0.18 \pm 0.12$  &  $0.85 \pm 0.37$ \\
 \hline
 \hline
 \multicolumn{5}{|c|} {
 {\boldmath \bf $\ecms=172$ GeV} \,\,\,\,\,\,(~$\int {\cal L} = 9.9$ pb$^{-1}$) } \\
   $1$  & $0.07 \pm 0.05$  & -  & $0.08 \pm 0.08$  &  $0.21 \pm 0.10$  \\
 \hline
 \hline
 \multicolumn{5}{|c|} {
 {\boldmath \bf $\ecms=183$ GeV} \,\,\,\,\,\,(~$\int {\cal L} = 50.0$ pb$^{-1}$) } \\
   $4$  & $0.08 \pm 0.06$ & $1.15 \pm 0.40$  & $0.34 \pm 0.13$  &  $1.67 \pm 0.74$ \\
 \hline
 \hline
 \multicolumn{5}{|c|} {
 {\boldmath \bf $\ecms=189$ GeV} \,\,\,\,\,\,(~$\int {\cal L} = 152.9$ pb$^{-1}$) } \\
   $8$  & $1.09 \pm 0.44$ & $3.50 \pm 1.25$  &  $1.21 \pm 0.42$ &  $5.85 \pm 1.39$ \\
 \hline
 \hline
 \multicolumn{5}{|c|} {
 {\boldmath \bf Sum of all centre-of-mass energies } } \\
   $13$   & $1.2 \pm 0.4$  & $5.4 \pm 1.4$  & $1.8 \pm 0.5$  &  $8.7 \pm 1.6 $  \\
 \hline
 \end{tabular}
 \end{center}
\caption[]{ Events remaining in the data and in the sum of the expected SM
    backgrounds after the OR of the selections applied in the search for nearly
    mass-degenerate charginos accompanied by high $p_T$ ISR photons. The
    breakdown of the three most important sources of background is given,
    together with the total background.
   }
 }
\label{tab:isrcand}
\end{table}

\begin{figure}[tbh]
\centerline{
\epsfxsize=11.cm\epsffile{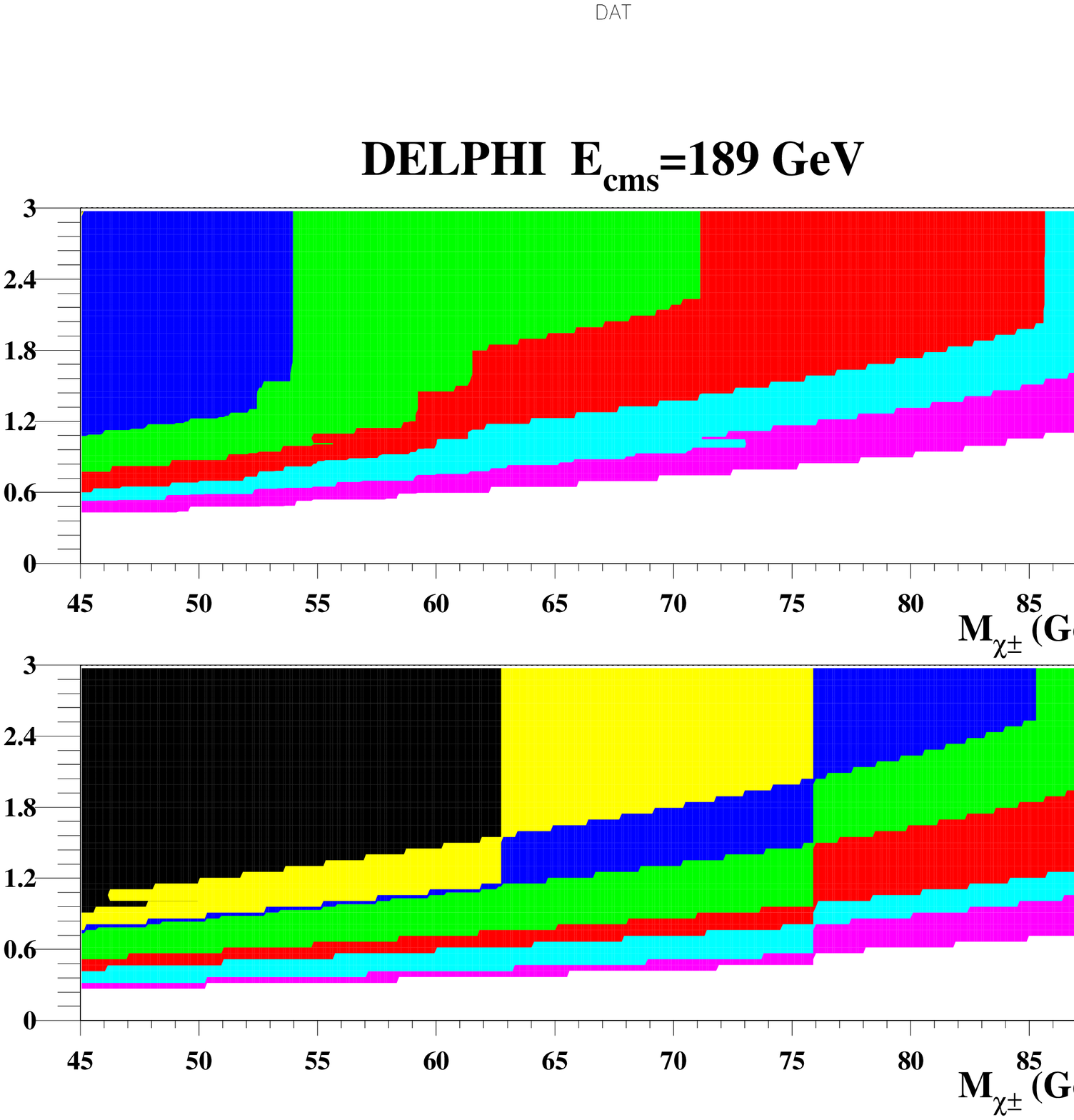} }
\caption[]{ SM MC (top) and data (bottom) remaining after the final selection in
  any point of the  plane ($\mcgup$,$\DMP$) considered in the search with the
  ISR photon at 189 GeV.
                     }
\label{fig:fondi}
\end{figure}

The selection cuts are different for different points of the plane 
($\mcgup$,$\DMP$).
Therefore, also the expected background content and the number of events
remaining in the data are different in any point of that plane. 
Figure~\ref{fig:fondi} shows the expected background (top) and data (bottom)
content in the different points of the plane, at 189 GeV. Similar figures have
been obtained for the other centre-of-mass energies.

After the selection, the data remaining are compatible with coming from 
background alone and there is no evidence of any significant excess above
the SM expectations. Then, the data collected at all LEP2 energies were
combined and used to set lower limits on the mass of the chargino in nearly
mass-degenerate scenarios. 

The procedure for combining results at the different energies was similar to
that of Ref.~\cite{paper214}. This procedure takes into account the effect on
the limit of the uncertainties on the signal efficiencies and on the background
content. The interpolation of the selection and trigger efficiencies between 
the points of the plane ($\mcgup$,$\DMP$) where a full simulation was produced,
was based on SUSYGEN samples produced only at the generator level.

The regions excluded with at least 95\%~CL by the search with the high $p_T$
ISR photon tag in the plane ($\mcgup$,$\DMP$) after such combination are shown
in figure~\ref{fig:limit}. Those limits were obtained by subtracting the
SM background included in the available simulated samples, possibly incomplete 
(see par.~\ref{sec:data}); for that reason the confidence levels obtained are
likely to be underestimated (i.e. the limits are conservative).
Since $\DMP=170$~\mevc2 is the smallest $\DMP$ fully simulated for the search
with the ISR tag, ISR efficiencies are supposed to vanish completely for all
mass differences smaller than 170~\mevc2

\section{Limit on the mass of nearly degenerate charginos}

\begin{figure}[tbh]
\centerline{
\epsfxsize=14.cm\epsffile{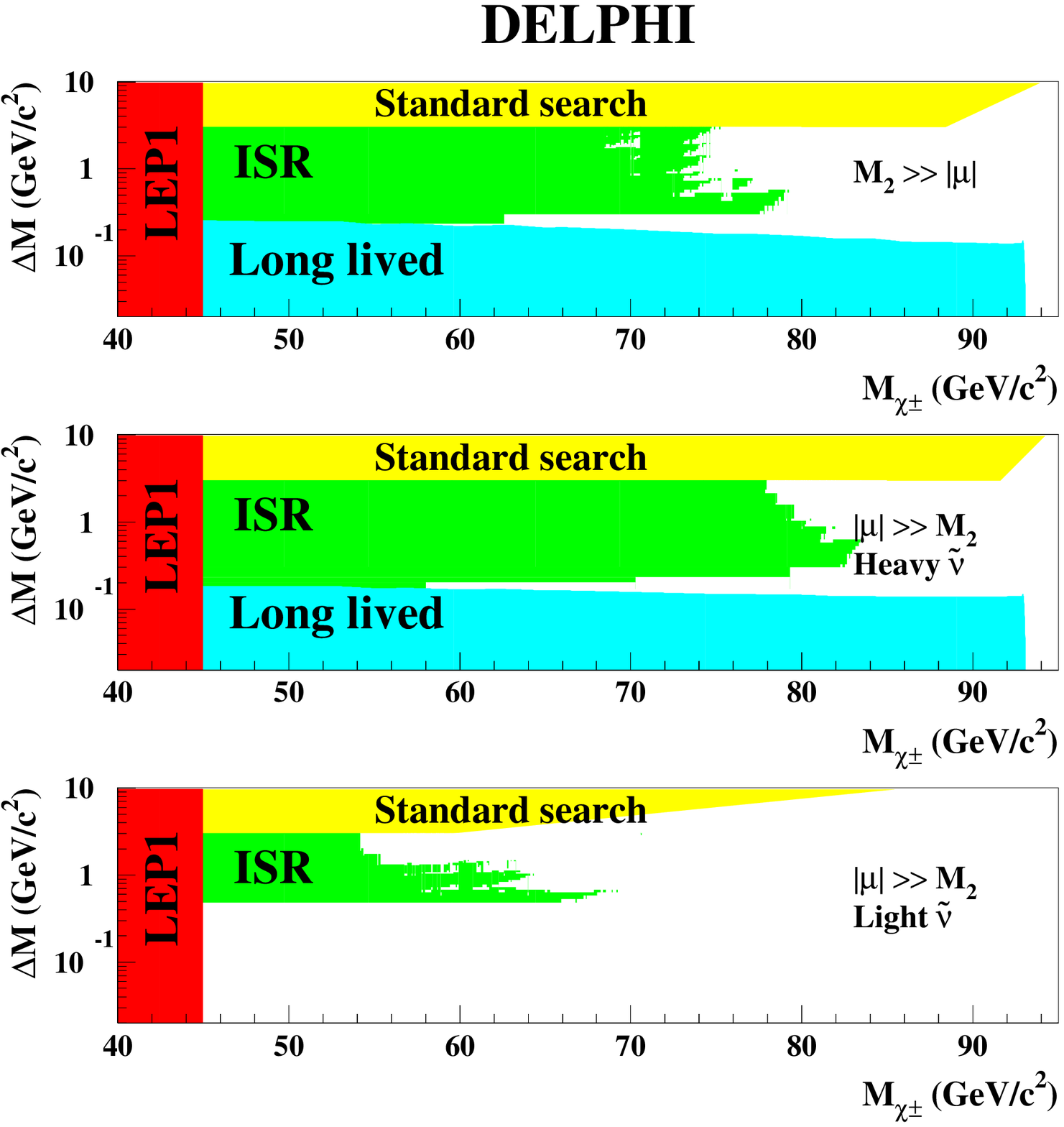} }
\caption[]{ Regions in the plane ($\mcgup$,$\DMP$) excluded by DELPHI with at
  least $95\%$~CL using separately the search for high $\DMP$ charginos, the
  search for soft particles accompanied by ISR and the search for long-lived
  charginos, in the three scenarios with near mass-degeneracy between the
  chargino and the lightest neutralino. Better limits in the region of overlap
  of the different search methods are obtained after combining them in logical
  OR (see text).
                     }
\label{fig:limit}
\end{figure}

The results of the searches for long-lived charginos and for soft particles
accompanied by a high $p_T$ photon at 189 GeV have been combined with the
results of the searches at lower energies to obtain the excluded regions in the
plane ($\mcgup$,$\DMP$) shown in figure~\ref{fig:limit}. The same figure shows,
for comparison, also the region excluded by the independent search in DELPHI
for charginos with larger $\DMP$ \cite{charpaper}.

By simply superimposing the regions excluded with the search for long-lived
charginos and the regions excluded with the search with the ISR photon tag,
these results permit to exclude $\mcgup < 55.6$~\gevc2 for any $\DMP$, in the
higgsino scenario and if $\msnu>\mcgup$. In the gaugino scenario with heavy 
$\snu$'s one can also exclude $\mcgup < 58.1$~\gevc2.
The narrow non-excluded bands between the exclusions given separately by the
searches for long lived charginos and charginos plus ISR can be covered when
combining in logical OR the two search methods. In that case the two limits
rise to 62.4~\gevc2 and 59.8~\gevc2, respectively. Given the efficiencies of
the two search methods (see figures \ref{fig:kinkeff} and \ref{fig:isreff}),
the result of the combined search is different from the separate ones only
in the narrow band where $\DMP>170$~\mevc2 and there is a significant number 
of chargino decays with decay length above 10~cm.

All the exclusion are with a confidence level of at least $95\%$.

These limits take into account a variation of $\tan \beta$ between $1$ and 
$50$, and a variation of $M_1$, $M_2$ and $\mu$ such that the mass difference
between the chargino and the neutralino remains below $3$~\gevc2 and 
$M_2 \le 2M_1 \le 10M_2$.

The sneutrino is always considered to be heavier than the chargino 
($\msnu>500$~\gevc2 in the second scenario).
% $\msnu>\mcgup+500$~\mevc2 for the
%long-lived charginos in the third scenario). 
If $\msnu \le \mcgup$, the limits derived in the searches for long-lived
charginos are not valid any more, even in the higgsino scenario (because the
two body decay $\cgup \to \snu l^+$ opens up for the gaugino component, which
is always present because of mixing). However, the search exploiting the ISR
tag remains sensitive even if $\msnu \le \mcgup$, with the relevant
$\DMP = \mcgpm - \msnu$. The corresponding mass limit is expected to extend to
slightly higher $\mcgpm$, as compared to what was found for heavier sneutrinos
(third plot of figure~\ref{fig:limit}),
at the smallest $\DMP$ studied with the ISR method. On the contrary, the limit
is expected to be somewhat reduced when approaching from below $\DMP=3$~\gevc2.
In both cases, the effect arises because of the larger mean energy of the 
visible products in the two-body chargino decay, which is expected to increase
the detectability at very low $\DMP$ but, at the contrary, tends to lower the
selection efficiency when the upper bound on the total visible energy applies.

As far as the masses of the scalar partners of the SM charged fermions are
concerned, they were supposed to be large enough in order to not modify
in a significant way the lifetimes and branching ratios used to obtain the
present results (in the higgsino scenario it is sufficient that 
$M_{\tilde f_i} > \mcgup$).

\section{Conclusions}

Charginos nearly mass-degenerate with the lightest neutralino were searched
for in DELPHI using the data collected at 189~GeV. 
Two different searches for long-lived charginos were complemented with a search
for nearly mass-degenerate charginos that exploits the tag of an ISR photon.
An improved selection was used for the search with the ISR photon tag, as
compared with the previous analysis. No evidence of a signal was found.
The results of the searches at 189~GeV were combined with those obtained
at lower centre-of-mass energies, where the old samples were re-analysed 
according to the new selection criteria, whenever different. 

The regions excluded with CL~$\ge 95\%$ in the space of SUSY parameters were
thus extended in all scenarios in which the chargino and the lightest neutralino
acquire similar masses. In particular, if all sfermions are heavy, a lower 
limit of $59.8$~\gevc2 on the mass of the chargino can be derived, 
independently on  $\DMP = \mcgup - \mcguz$ and for any field composition of 
the chargino. If the MSSM gaugino masses unify at the GUT scale, the charginos
can only be almost pure higgsinos if nearly mass degenerate with the lightest
neutralino. In this case, the CL~$\ge 95\%$ $\DMP$ independent lower limit on
the mass of the chargino is $62.4$~\gevc2, and this limit is valid whenever
$M_{\tilde f_i} > \mcgup$, where $\tilde f_i$ represents any scalar partner
of the SM fermions.

%         Modified on 04-06-1999 by dimartino
%-------------------------------------------------------------------
\subsection*{Acknowledgements}
\vskip 3 mm
We thank M. Drees for valuable comments and suggestions.\\
 We are also greatly indebted to our technical
collaborators, to the members of the CERN-SL Division for the excellent
performance of the LEP collider, and to the funding agencies for their
support in building and operating the DELPHI detector.\\
We acknowledge in particular the support of \\
Austrian Federal Ministry of Science and Traffics, GZ 616.364/2-III/2a/98, \\
FNRS--FWO, Belgium,  \\
FINEP, CNPq, CAPES, FUJB and FAPERJ, Brazil, \\
Czech Ministry of Industry and Trade, GA CR 202/96/0450 and GA AVCR A1010521,\\
Danish Natural Research Council, \\
Commission of the European Communities (DG XII), \\
Direction des Sciences de la Mati$\grave{\mbox{\rm e}}$re, CEA, France, \\
Bundesministerium f$\ddot{\mbox{\rm u}}$r Bildung, Wissenschaft, Forschung
und Technologie, Germany,\\
General Secretariat for Research and Technology, Greece, \\
National Science Foundation (NWO) and Foundation for Research on Matter (FOM),
The Netherlands, \\
Norwegian Research Council,  \\
State Committee for Scientific Research, Poland, 2P03B06015, 2P03B1116 and
SPUB/P03/178/98, \\
JNICT--Junta Nacional de Investiga\c{c}\~{a}o Cient\'{\i}fica
e Tecnol$\acute{\mbox{\rm o}}$gica, Portugal, \\
Vedecka grantova agentura MS SR, Slovakia, Nr. 95/5195/134, \\
Ministry of Science and Technology of the Republic of Slovenia, \\
CICYT, Spain, AEN96--1661 and AEN96-1681,  \\
The Swedish Natural Science Research Council,      \\
Particle Physics and Astronomy Research Council, UK, \\
Department of Energy, USA, DE--FG02--94ER40817. \\
%=========================================================================%

\end{document}